\documentclass[twocolumn,showpacs,preprintnumbers,amsmath,amssymb]{revtex4-1}
\usepackage{graphicx}
\usepackage{dcolumn} 
\usepackage{bm}      
\usepackage{xcolor}      
\usepackage[mathscr,mathcal]{euscript}

\begin{document}

\title{Plasmon modes in the magnetically doped Single Layer and Multi layers of Helical Metals}

\author{Fei Ye$^{1}$ and Chaoxing Liu$^{2}$}

\affiliation{$^{1}$Department of Physics, South University of Science
  and Technology of China, Shenzhen 518055, China}
\email{ye.f@sustc.edu.cn}
\affiliation{$^{2}$Department of Physics, The Pennsylvania State University, University Park, Pennsylvania 16802-6300}
\email{cxl56@psu.edu}
\date{\today}

\begin{abstract}
  We study the plasmon excitations and the electromagnetic response of
  the magnetically doped single layer and multilayer of ``helical
  metals'', which emerge at the surfaces of topological insulators. For
  the single layer case, we find a ``spin-plasmon'' mode with the
  rotating spin texture due to the combination of the spin-momentum
  locking of ``helical metals'' and the Hall response from
  magnetization. For the multilayers case, we investigate the
  electromagnetic response due to the plasmon excitations, including the
  Faraday rotation for the light propagating normal to the helical metal
  layers and an additional optical mode with the frequency within the
  conventional plasmon gap for the light propagating along the helical
  metal layers.
\end{abstract}
\maketitle

\section{Introduction}
\label{sec:introduction}
Topological insulators (TIs)\cite{qi2010a,moore2009,hasan2010,qi2011},
as a class of materials with insulating bulk states, but conducting
surface states, have been theoretically
proposed\cite{kane2005a,kane2005b,bernevig2006a,bernevig2006,fu2007,qi2008b,moore2007,fu2007b,zhang2009}
and experimentally realized both in two dimensions, such as HgTe quantum
wells\cite{konig2007}, and three dimensions\cite{hsieh2008}, such as
Bi$_2$Se$_3$ family of materials\cite{xia2009,chen2009}. For three
dimensional (3D) TIs, for example, Bi$_2$Se$_3$, the surface state
consisting of a single two dimensional (2D) Dirac cone with the gapless
Dirac point protected by time reversal symmetry since the two components
of the Dirac cone are Kramers' partners. For simplicity, we can regard
these two components as ``spin''. The 2D surface state exhibits a novel
``spin-momentum locking''\cite{wu2006,fu2007,zhang2009,liu2010}, in the
sense that the ``spin'' is locked and points perpendicularly to the
momentum, forming a left-hand helical texture in the momentum
space. Therefore, the surface state of TIs is also dubbed ``helical
metals'' (HMs)\cite{wu2006,raghu2010}. For a finite Fermi energy $E_f$,
the HM possesses a plasmon excitation similar to graphene with the
dispersion $\sim k^{1/2}$ in the long wavelength limit (the momentum
$k\rightarrow 0$). But different from graphene, this plasmon excitation
is coupled to the spin wave due to the helical nature, leading to a
novel ``spin-plasmon'' mode\cite{raghu2010}.

Another intriguing feature of a HM is the Hall response due to the lift of 
the degeneracy at the Dirac point by the breaking of time reversal
symmetry, when the surface of TIs is coated by a ferromagnetic layer, or
doped with magnetic atoms, which provide magnetization normal to the surface. 
If the Fermi surface lies inside the
magneto-bandgap, the Hall conductance is half quantized as $e^2/(2h)$,
known as ``half quantum Hall
effect''\cite{redlich1984b,hasan2010,qi2011,qi2008b,fu2007}. In case of a finite Fermi surface, the Hall conductance
is no longer quantized, but still non-zero. For a non-zero Hall
response, the longitudinal and transverse electromagnetic
self-excitations are coupled, forming a magnetoplasmon, similar to that
in the conventional metals in a strong magnetic
field\cite{abrikosov88}. It is interesting to ask what is the
behavior of the ``spin-plasmon'' mode in the magnetically doped HMs.

In this paper, we investigate the interaction between electromagnetic
fields and HMs on the surface of a TI with magnetic doping, focusing on
the plasmon excitation. In Sec. II, we derive the correlation function
for a 2D single layer of HMs and discuss how the ``spin-plasmon'' mode
is affected by the magnetic doping. The electromagnetic waves exist in
three dimensions and can not interact strongly with the 2D HMs, so we
consider a model with multilayers of HMs, with an alternating stacking
of TIs and normal insulators. Based on the dielectric function of this
model, we discuss the propagating modes of the electromagnetic waves in
HMs, in Sec. III. The conclusion is drawn in Sec. IV.

\section{A single-layer of Helical Metal}
\subsection{Correlation functions}
We start from the calculation of the correlation function of a single
layer of HMs on the surface($xy$ plane) of TIs with magnetic doping.
The single particle Hamiltonian\cite{liu2010,fu2009c} for this system is
written by
\begin{eqnarray}
\label{eq:3}
\mathcal{H}_{\vec{k}}=
\hbar v_f(\vec{k}\times\vec{\sigma})_z+m\sigma_z-E_f \sigma_0
\end{eqnarray}
where $\{\sigma_x,\sigma_y,\sigma_z \}$ are the Pauli matrices,
$\sigma_0$ is the $2\times2$ identity matrix and $v_f$ and $E_f$ are the
Fermi velocity and Fermi energy respectively. The first term on the
right hand side is the kinetic term, taking the form of Rashba
spin-orbit coupling, while the second term gives the Zeeman type of
coupling due to magnetization. Here we only consider the magnetization
along $z$-axis for simplicity and the Zeeman coupling strength $m$ gives
the gap of the HMs. The inplane current operator reads
\begin{eqnarray}
\label{eq:4}
\vec{\gamma}
 = e v_f (\hat{z}\times \vec{\sigma}). 
\end{eqnarray}
According to the linear response theory, the current-current correlation
functions read 
\begin{eqnarray}
\label{eq:5}
&&\chi_{ab}(\vec{k},i\omega_n) \nonumber\\
&=& 
\frac{1}{\beta \mathcal{A}}\sum_{\vec{q},i\nu_n} 
\text{tr} \left[ G_0(\vec{q},i\nu_n)\gamma_a
G_0(\vec{k}+\vec{q},i\omega_n+i\nu_n)\gamma_b \right]
\end{eqnarray}
where $\mathcal{A}$ is the area of the surface, $a,b$ are $x$ or $y$,
and $\beta$ is the inverse temperature. $G_0(\vec{k},i\omega_n)$ is the
single particle Matsubara's function
\begin{eqnarray}
\label{eq:1}
G_0(\vec{k},i\omega_n) = (i\hbar\omega_n-\mathcal{H}_{\vec{k}})^{-1}, 
\end{eqnarray}
with the Matsubara frequency $\omega_n$. In the long wavelength limit
$k\rightarrow0$, the rotation symmetry is restored and the correlation
function is restricted to the following form
\begin{eqnarray}
\label{eq:SL_chi}
\chi(\vec{k},\omega)=
\chi_0(\omega)- \chi_H(\omega)\sigma_y+o(k^2)  
\end{eqnarray}
where we only keep the momentum independent term for both the diagonal
part $\chi_{0}(\omega)$ and the Hall response $\chi_{H}(\omega)$, which
are calculated straightforwardly as
\begin{eqnarray}
\label{eq:2}
&&\chi_{0}(\omega) =\frac{e^2E_f}{4\pi \hbar^2} \left[ 
 1- \frac{x^2+(\frac{m}{E_f})^2}{2x}
  f(x)  \right],
  \label{eq:SL_chi0}\\
&&\chi_H(\omega) = \frac{e^2m}{4\pi \hbar^2}f(x),
\label{eq:SL_chiH}\\
&&f(x)\equiv\ln \left| \frac{1+x}{1-x} \right| +i \pi\theta(x-1)
\end{eqnarray}
with $x\equiv \hbar\omega/(2E_f)$. The step function $\theta(x)$ in the
imaginary part of $f(x)$ indicates the appearance of the particle-hole
excitation for the frequency $\hbar\omega>2E_f$ in the $k=0$ limit.

Eqs. (\ref{eq:SL_chi}), (\ref{eq:SL_chi0}) and (\ref{eq:SL_chiH}) are
the main results of this section. These expressions can recover early
results in some limits. For example, without magnetization ($m=0$), the
system has a longitudinal plasmon excitation with dispersion
proportional to $k^{1/2}$ determined by Eq.~\eqref{eq:SL_chi0}, the same
to the results given in \cite{raghu2010}(see next subsection). The Hall
conductance in the low frequency limit ($\omega\rightarrow 0$) can be
determined by Eq.~\eqref{eq:SL_chiH}
\begin{eqnarray}
\label{eq:7}
\sigma_H = \lim_{\omega\rightarrow0} \frac{i\chi_{xy}(\omega)}{\omega}
= -\frac{me^2}{4\pi\hbar E_f}
\end{eqnarray}
As $E_f\rightarrow m$, the system becomes insulating, and the Hall
conductance tends to be $e^2/(2h)$, yielding the ``half quantum
Hall effect''\cite{redlich1984b,hasan2010,qi2011,qi2008b,fu2007}.

\subsection{``Spin-plasmon'' modes}
\label{sec:spin-plasmon-modes}
Now let's consider the plasmon excitation in the presence of Hall
response from Eqs.~\eqref{eq:SL_chi0} and \eqref{eq:SL_chiH}. Due to the
existence of the Hall response, the plasmon excitation is not purely
``longitudinal'' anymore, but turns out to be a mixture of both the
longitudinal and transverse modes. To calculate these hybrid
excitations, we first note that for charges and currents distributed in
a 2D sheet at $z=0$, the electric and magnetic fields \emph{in the
  plane} can be calculated through the Coulomb's law and Biot-Savart law
\begin{eqnarray}
\label{eq:31}
\vec{E}(\vec{r},t) &=& \int d^2\vec{r}'
\frac{\rho(\vec{r}',t)(\vec{r}-\vec{r}')}{\epsilon_d|\vec{r}-\vec{r}'|^{3/2} }
\\
\label{eq:32}
B_z(\vec{r},t) &=& \frac{\mu_d}{c}
 \int d^2\vec{r}'
\frac{[\vec{J}(\vec{r}',t)\times(\vec{r}-\vec{r}')]_z}{|\vec{r}-\vec{r}'|^{3/2} } 
\end{eqnarray}
where $\vec{r}$, $\vec{r}'$, $\vec{J}$ and $\vec{E}$ are all
\emph{inplane} vectors, and $\mu_d$ and $\epsilon_d$ are the
permeability and permittivity of the surroundings respectively. Here,
the magnetic field in the plane $z=0$ has only $z$ component generated
by the inplane current. In Eqs.~\eqref{eq:31} and \eqref{eq:32} we
neglect the effect of retardation\cite{fetter1973} which does not
affect the key results discussed below and will be considered in the
appendix.

Without loss of generality, we assume the wave is propagating along $y$
axis in the plane, so that the subscripts $x$ and $y$ correspond to the
transverse and longitudinal component, respectively. In fact
Eq.~\eqref{eq:31} gives the longitudinal electric field, while
Eq.~\eqref{eq:32} leads to the transverse one according to the Faraday's
law $(\vec{\nabla}\times \vec{E})_z=-c^{-1}\partial_tB_z$. Then, after the
Fourier transformation of Eq.~\eqref{eq:31} and Eq.~\eqref{eq:32}, one
obtains the inplane electric fields in the plane,
\begin{eqnarray}
\label{eq:21}
E_x(\vec{k},\omega) &=&
\frac{2\pi i\mu_d\omega}{c^2k} J_x(\vec{k},\omega) \\
\label{eq:22}
E_y(\vec{k},\omega)&=&
-\frac{2\pi ik}{\epsilon_d \omega} J_y(\vec{k},\omega),
\end{eqnarray}
where we have applied the continuity condition
$\omega\rho(\vec{k},\omega) = \vec{k}\cdot \vec{J}(\vec{k},\omega)$, to
replace the charge density in Eq.~\eqref{eq:31} by the longitudinal
current.  Combined with Ohm's law $\vec{J}=\sigma(\omega)
\vec{E}=i(\chi/\omega)\vec{E}$, Eqs.~\eqref{eq:21} and \eqref{eq:22}
give rise to the self-consistent equation for electric fields
\begin{eqnarray}
\label{eq:13}
\begin{pmatrix}
\chi_0(\omega) + \frac{c^2k}{2\pi \mu_d} & i\chi_H(\omega) \\
-i\chi_H(\omega) & \chi_0(\omega)- \frac{\omega^2\epsilon_d}{2\pi k}
\end{pmatrix} \begin{pmatrix}
 E_x \\ E_y
\end{pmatrix}
=0.
\end{eqnarray}
Obviously, without magnetization, there is no transverse solution of the
above equation, i.e., $E_x=0$. Only the longitudinal excitation exists
with dispersion determined by
\begin{eqnarray}
\label{eq:6}
k = \frac{\epsilon_d\omega^2}{2\pi \chi_0(\omega)}.
\end{eqnarray}
In the limit $\omega\ll E_f$, one immediately gets the plasmon frequency
$\omega=\alpha k^{1/2}$ with the coefficient $\alpha=|ev_fk_f|/ (
\sqrt{2\epsilon_dE_f})$, recovering the results in \cite{raghu2010}.
Here $k_f$ is the Fermi momentum, and other quantities are defined as
before. It is emphasized that the plasmon dispersion thus obtained has
the divergence of group velocity as $k\rightarrow0$, which is unphysical
due to the neglecting of the effect of retardation which is rather
strong for small $k$ (see the appendix)\cite{fetter1973}.

In the presence of magnetization, the dispersion of the collective
mode in HMs is determined by vanishing the determinant of the
matrix in Eq.~\eqref{eq:13}, which in the small $k$ limit has the
following form
\begin{eqnarray}
\label{eq:19}
\omega(k) = \left[ \frac{\left( \frac{2\pi\sigma_H}{\alpha\epsilon_d}
    \right)^2}{ \alpha^2+\frac{c^2k}{\epsilon_d\mu_d}} +
  \frac{1}{\alpha^2k} \right]^{-1/2}. 
\end{eqnarray}
In this case both $E_x$ and $E_y$ are nonzero, and the ratio between
them is determined by
\begin{eqnarray}
\label{eq:18}
\frac{E_y}{E_x} = i  \frac{\chi_0(\omega)+
  \frac{c^2k}{2\pi\mu_d}}{\chi_H(\omega)} 
=i\frac{\chi_H(\omega)}{\chi_0(\omega)-\frac{\omega^2\epsilon_d}{2\pi k}}.
\end{eqnarray}
It is a pure imaginary number, therefore, the electric field is not
simply longitudinal but \emph{elliptically} polarized in the plane of
HMs for a given $\vec{k}$. 

\begin{figure}[htbp]
\centerline{\includegraphics[width=5cm]{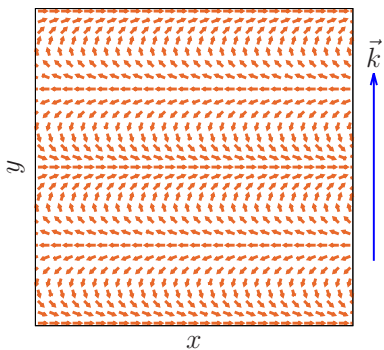}}
\caption[]{\label{fig:1} An illustration of spin orientation indicated
  by the arrows in the $xy$ plane with $S_y/(iS_x)$ positive, and the
  wavevector $\vec{k}$ is along $y$ axis.  }
\end{figure}

Due to the spin-orbit coupling, the collective mode in HMs can also be
viewed as the spin plasmon\cite{raghu2010}.  Unlike the nonmagnetized
case in Ref.\onlinecite{raghu2010} where a simple inplane linear
polarized spin density fluctuation is found, the spin density as a
vector field in the magnetized case is rotating elliptically in the $xy$
plane, and the ratio between the spin densities $S_x$ and $S_y$
satisfies
\begin{eqnarray}
\label{eq:20}
\frac{S_x}{S_y} = 
-\frac{J_y}{J_x} = 
\frac{\omega^2\epsilon_d\mu_d}{c^2k^2} \frac{E_y}{E_x}
\end{eqnarray}
The spin texture $\vec{S}(\vec{r})$ in the xy plane is shown in Fig.~\ref{fig:1}

\section{Multilayers of helical metals} 
\label{sec:diel-mult-helic}
\subsection{Dielectric functions and plasmon modes}
\begin{figure}[htbp]
\centerline{\includegraphics[width=5cm]{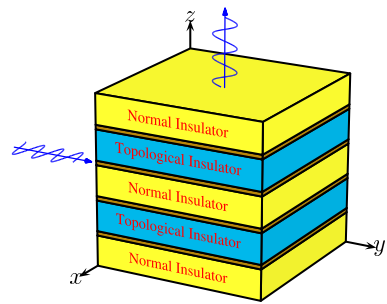}}
\caption[]{\label{fig:2} Illustration of layered HMs by
  stacking topological insulator and normal insulator alternatively. In
  the text, we study two basic situations of incidence, either along $y$
  or $z$ axis as indicated by the blue lines. }
\end{figure}
We now consider the response of the macroscopic electromagnetic fields
for layered HMs by stacking alternating layers of the TIs and normal
insulators as shown in Fig.~\ref{fig:2}, with the magnetization
perpendicular to HM layers.

The planes of HMs are equal spacing with the distance $l_d$ between
adjacent layers. The local current density can be written as
$\vec{J}_{loc}(\vec{k},z,\omega)=\sum_n
\vec{J}_n(\vec{k},\omega)\delta(z-nl_d)$, which is related to the local
electric field by the Ohm's law $\vec{J}_{n}(\vec{k},\omega)
=(i\chi/\omega) E_{loc}(\vec{k},z=nl_d,\omega)$ where the momentum
$\vec{k}$ lies in the xy plane.  The macroscopic current density and
electric field in the $x,y$ direction can be obtained by averaging over
the local current and local electric field, respectively,
\begin{eqnarray}
\label{eq:30}
&&\vec{\mathcal{J}}(\vec{k},z,\omega) = \int dz' \xi(z-z')
\vec{J}_{loc}(\vec{k},z',\omega) \nonumber\\
&&\vec{\mathcal{E}}(\vec{k},z,\omega) = \int dz' \xi(z-z') \vec{E}_{loc}(\vec{k},z',\omega)
\end{eqnarray}
with a test function $\xi(z-z')$ following the standard average method
for macroscopic electrodynamics\cite{jackson1999}.  We assume $l_d$ to
be much smaller than the wavelength of the incident wave, but $l_d$ is
still large enough so that no tunnelings occur between two adjacent
layers. 

Under these assumptions and using the definition of macroscopic
displacement current $\partial_t\mathcal{D}=\partial_t\mathcal{E}+4\pi
\mathcal{J}$ we obtain the dielectric function
\begin{eqnarray}
\label{eq:8}
\epsilon(\omega) = \begin{pmatrix}
 \epsilon_d - \frac{4\pi\chi_0(\omega)}{l_d\omega^2} &
-i \frac{4\pi\chi_H(\omega)}{ l_d\omega^2} & 0 \\
i\frac{4\pi\chi_H(\omega)}{ l_d\omega^2} &
\epsilon_d-\frac{4\pi\chi_0(\omega)}{l_d\omega^2} & 0  \\
0&0&\epsilon_d
\end{pmatrix},
\end{eqnarray}
where $\chi_0$ and $\chi_H$ are defined in Eq.~\eqref{eq:SL_chi0} and
\eqref{eq:SL_chiH}. Eq.~\eqref{eq:8} immediately leads to two branches
of \emph{inplane} collective modes determined by
$\det[\epsilon(\omega)]=0$.  In the nonmagnetic case without Hall
response, one obtains the familiar plasmon excitations in the normal
metal in the limit $\omega\ll E_f$
\begin{eqnarray}
\label{eq:9}
\omega_{p0}^2 = \frac{e^2v_f^2k_f^2}{l_d\epsilon_dE_f}=
\frac{e^2v_f^2n}{\epsilon_dE_f} 
\end{eqnarray}
where $n\equiv k_f^2/(4\pi l_d)$ is the 3D electron density. Since
electrons can not move in the z direction, the $E_z$ component of
electromagnetic waves does not couple to the HMs. Eqs.~\eqref{eq:8} and
\eqref{eq:9} are obtained under the assumption that the electrons in
different layers within the range of wavelength are moving in phase,
since the macroscopic fields are the simple average of local ones, which
is consistent with our purpose of discussing optical properties for long
wavelength of electromagnetic wave. There is also another type of
excitation in which the electrons in the adjacent layers moves
out-of-phase\cite{fetter1973} and we will not consider it here.  In case
$m\ne0$, the frequencies of the two branches of plasma split and are
determined by
\begin{eqnarray}
\label{eq:11}
\omega^2_{p\pm} = \frac{4\pi[\chi_0(\omega_{p\pm})\pm\chi_H(\omega_{p\pm})]}{l_d\epsilon_d}.
\end{eqnarray}
The electric fields for these two plasmon modes are not oscillating
linearly, but rotating clockwise and anticlockwise in the plane,
respectively.  These \emph{inplane} plasmon excitations affect the
electromagnetic wave in the bulk with \emph{inplane} polarized electric
fields as shown in the next section.

\subsection{Electromagnetic response}
\label{sec:optic-prop-layer}
Given the wavevector $\vec{k}$ in normal metal, one can
distinguish the longitudinal plasmon mode ($\vec{E}\parallel \vec{k}$)
from the transverse optical mode($\vec{E}\perp \vec{k}$), both of which
are determined by solving the Maxwell equation
\begin{eqnarray}
\label{eq:24}
 \left[ \frac{\omega^2}{c^2}(\mu\epsilon)_{ij} - k^2\delta_{ij} +
   k_ik_j \right] E_j(\vec{k},\omega) = 0 .
\end{eqnarray}
However in the layered HMs with permittivity given in Eq.~\eqref{eq:8},
the Hall response might mix the longitudinal and transverse modes,
giving rise to the unusual hybrid optical modes.  Since we are only
interested in the influence of the collective excitations on the
propagating electromagnetic wave, we consider two different situations
with the inplane polarized electric field, one of which is propagating
along $z$ axis, and the other is propagating in the $xy$ plane. For the
electromagnetic wave with the out-of-plane polarized electric field, it
is not affected by the inplane collective mode, thus will propagate in a
similar way as that in a normal insulator.

\subsubsection{Propagating along $z$ axis}
\label{sec:propagating-along-z}
\begin{figure}[htbp]
  \centerline{\includegraphics[height=3.8cm]{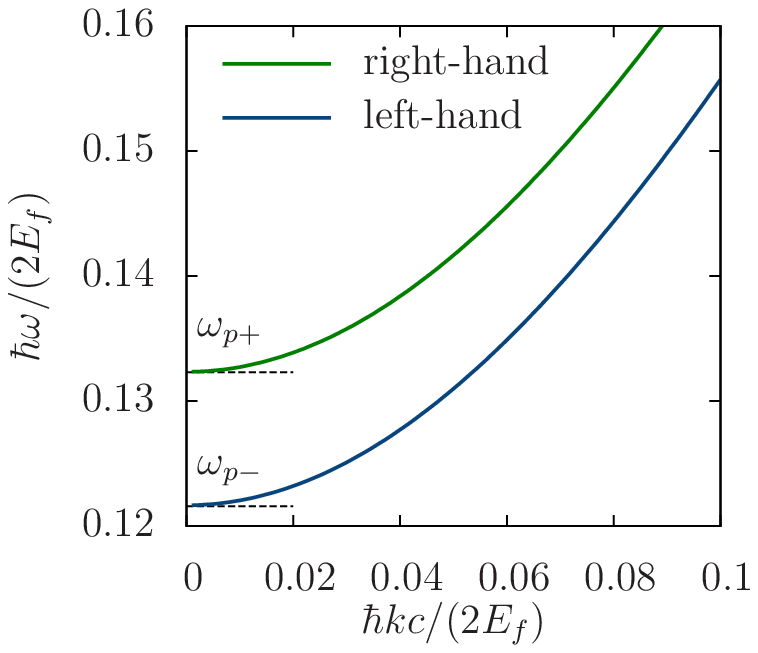}\includegraphics[height=3.8cm]{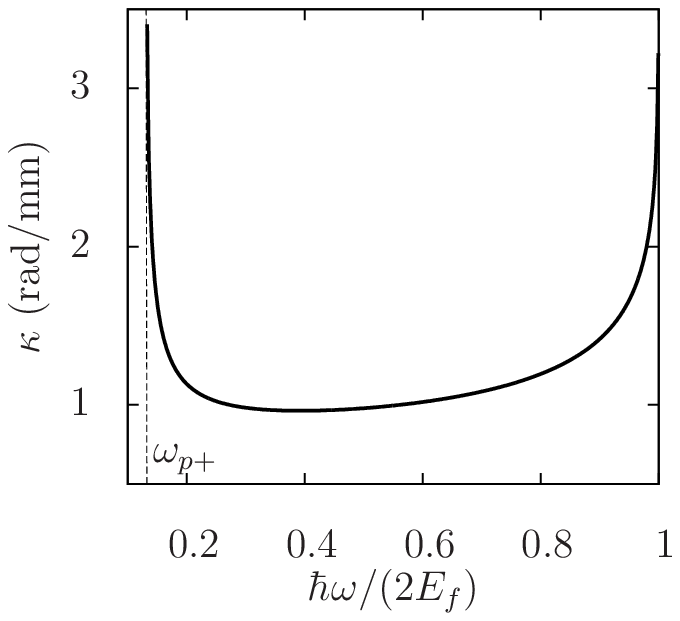}}
  \caption[]{\label{fig:3} Left panel: the dispersion for circularly
    polarized light propagating along $z$ axis, where we take
    $E_f=0.2eV$, $l_d=1000$\AA, $m=0.3E_f$ and $\epsilon_d=\mu_d=1$. The
    dispersions for left-hand(blue line) and right-hand(green line)
    polarization split.  Right panel: the angle per unit length $\kappa$
    (rad/mm) is plotted as a function of $\omega$.  }
\end{figure}

In this case we can take $k_x=k_y=0$, $k_z=k$, and $E_z=0$, and the
eigenmodes of an electromagnetic wave are simply the circularly
polarized wave, $E_y = \mp iE_x$, with the minus sign for the
\emph{left-hand} and the plus sign for the \emph{right-hand}
polarization, respectively. Both modes are transverse, i.e.,
$\vec{E}\perp \vec{k}$, as in the normal case, but the corresponding
dispersions are split as
\begin{eqnarray}
\label{eq:14}
\omega^2_{\pm} =
\frac{4\pi[\chi_0(\omega_{\pm})\pm\chi_H(\omega_{\pm})]}{\epsilon_dl_d}+ 
\frac{k^2c^2}{\epsilon_d \mu_d} .
\end{eqnarray}
due to the Hall response, as plotted in Fig.~\ref{fig:3}.  There are
threshhold frequencies at $k=0$ for the left-hand and right-hand modes
as seen in Fig.~\ref{fig:3}, which are $\omega_{p,\pm}$ given by
Eq.~\eqref{eq:11}. In our parameter setting, we have
$\omega_{p+}>\omega_{p-}$. Below $\omega_{p-}$, no electromagnetic wave
can transmit. If $\omega_{p-}<\omega<\omega_{p+}$, only the
\emph{right-hand} polarized one is allowed to transmit.
If $\omega>\omega_{p+}$, the Faraday rotation takes place. The Faraday rotation angle
can be calculated straightforwardly as following\cite{landau60}
\begin{eqnarray}
\label{eq:41}
\kappa(\omega)= \frac{k_{-}(\omega)-k_{+}(\omega)}{2} ,
\end{eqnarray}
with
\begin{eqnarray}
\label{eq:42}
k_{\pm}(\omega) = \frac{\omega \sqrt{\mu_d\epsilon_d}}{c}
\sqrt{1- \frac{4\pi[\chi_0(\omega)\pm\chi_H(\omega)]}{\epsilon_dl_d\omega^2}}.
\end{eqnarray}
We plot $\kappa$ as a function of $\omega$ in the right panel of
Fig.~\ref{fig:3}, where one find $\kappa$ reaches the maximum as
$\omega\rightarrow \omega_{p+}$ or $\omega\rightarrow 2E_f/\hbar$.  The
giant Faraday rotation at $\omega_{p+}$ is due to the small velocity of
the right-hand polarized light at its plasma frequency, while the rapid
increase of the Faraday rotation angle around $\hbar\omega/(2E_f)\sim 1$
is because the real excitation from the lower Dirac cone to the unoccupied
upper Dirac cone above Fermi energy starts to happen. The Faraday
rotation effect of the magnetized surface states of TIs has been
carefully investigated for the insulating
phase\cite{maciejko2010,tse2010}, and here we extend the discussion to
the metallic phase, which is more relevant to the present experimental
situation\cite{shuvaev2011}.

\subsubsection{Propagating parallel to $xy$ plane}
\label{sec:prop-parall-xy}
\begin{figure}[htbp]
\centerline{\includegraphics[width=6cm]{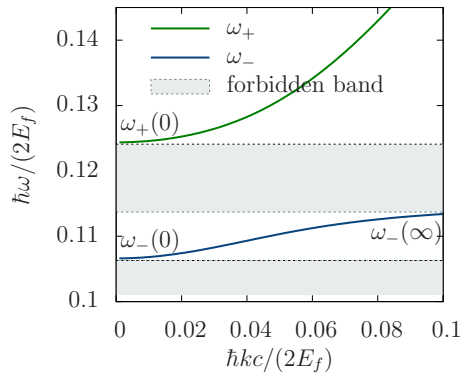}}
\caption[]{\label{fig:4}The dispersion of the two modes of
  non-transverse wave with $E_f=0.2eV$, $m=0.5E_f$, $l_d=1000$\AA, and
  $\epsilon_d=\mu_d=1$. }
\end{figure}

In this case, without loss of generality, we can take the incidence
along $y$-axis, i.e., $k_x=k_z=0$, $k_y=k$, and $E_z=0$. The frequencies
of the bulk excitations are determined by solving the following equation
\begin{eqnarray}
\label{eq:16}
\begin{pmatrix}
  1- \frac{4\pi\chi_0}{\epsilon_dl_d\omega^2}- \frac{k^2c^2}{\epsilon_d
    \mu_d\omega^2} &
 - i\frac{4\pi\chi_H}{\epsilon_dl_d\omega^2} \\
   i\frac{4\pi\chi_H}{\epsilon_dl_d\omega^2} & 1-
  \frac{4\pi\chi_0}{\epsilon_dl_d\omega^2}
\end{pmatrix} \begin{pmatrix}
 E_x \\ E_y
\end{pmatrix} =0 \;.
\end{eqnarray}
Obviously, if there is no magnetization, i.e., $\chi_H=0$, the two
eigenmodes are decoupled. One of them is the longitudinal plasmon
excitation, and the other is the transverse mode which couples to the
electromagnetic wave and determines the electromagnetic response of the
medium.  However in the presence of magnetization, the Hall response
couples these two excitations, leading to two hybridized modes in the
bulk.  The dispersions of the two modes are plotted in Fig.~\ref{fig:4},
where we denote the upper branch with $\omega_{+}(k)$ and the lower one
with $\omega_{-}(k)$. The lower branch, which originates from the
longitudinal plasmon in the nonmagnetic case, has a dispersion and
affects electromagnetic waves as well due to the mixing with the
transverse mode by the Hall response.
In Fig.~\ref{fig:4}, one finds two forbidden bands for the transmission
of the electromagnetic wave, which fall into the regimes
$[0,\omega_{-}(0)]$ and $[\omega_{-}(\infty),\omega_{+}(0)]$, where
$\omega_{-}(\infty)$ can be identified with the plasmon frequency
$\omega_{p0}$ given by Eq.~\eqref{eq:9} in the nonmagnetic case, while
$\omega_{\pm}(0)$ can be identified with $\omega_{p\pm}$ as given in
Eq.~\eqref{eq:11}. In the small $\omega$ limit, these threshold
frequencies have the following form
\begin{eqnarray}
\label{eq:35}
\omega_{\pm}(0) = \sqrt{\left( \frac{2\pi\sigma_H}{l_d\epsilon_d}
  \right)^2 + \omega_{p,0}^2} \pm \frac{2\pi|\sigma_H|}{l_d\epsilon_d}. 
\end{eqnarray}

\begin{figure}[htbp]
\centerline{\includegraphics[height=3.8cm]{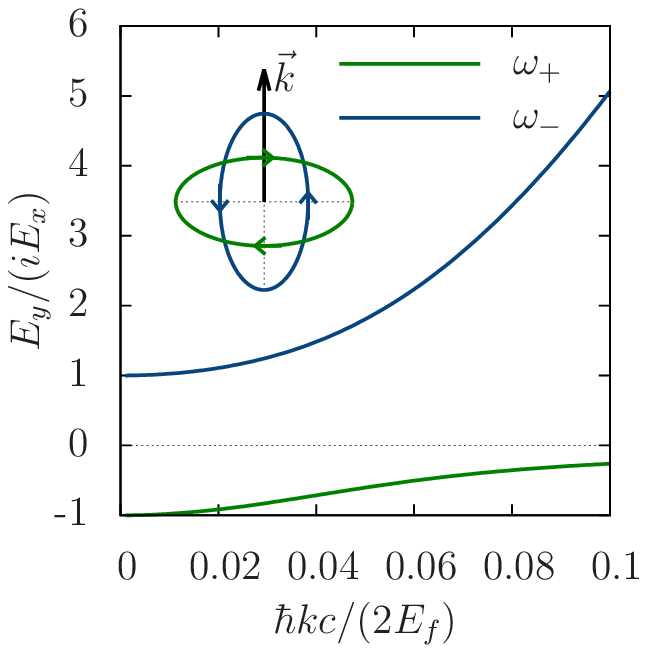}%
\includegraphics[height=3.8cm]{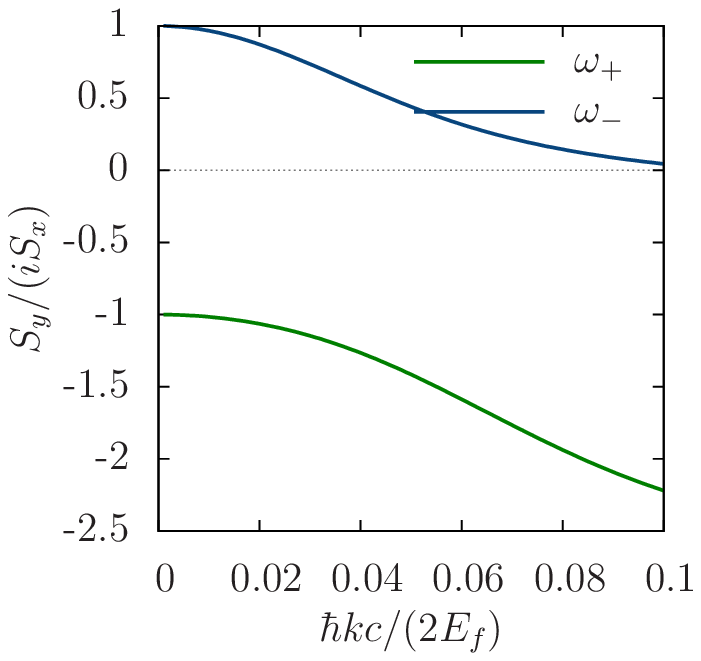}
}
\caption[]{\label{fig:5} The ratio $E_y/(iE_x)$ (left panel) and
  $S_y/(iS_x)$(right panel) of the inplane elliptic polarization of
  electric fields of the propagating wave with $E_f=0.2eV$,
  $l_d=1000$\AA, $m=0.3E_f$ and $\epsilon_d=\mu_d=1$. The wavevector
  $\vec{k}$ is along the $y$ axis. The inset of left panel gives the two
  inplane elliptical rotation modes of the electric fields. }
\end{figure}

Similar to the single layer case, the electric field is neither
perpendicular nor parallel to the wave vectors, instead, it is
\emph{elliptically} polarized in the $xy$ plane, as shown in the inset
of the left panel of Fig.~\ref{fig:5}, where we also plot $E_y/(iE_x)$
as a function of $\vec{k}$ which is determined by Eq.~\eqref{eq:16}
\begin{eqnarray}
\label{eq:17}
\frac{E_y}{iE_x} =  \frac{4\pi\chi_H}{4\pi\chi_0-\epsilon_dl_d\omega^2} .
\end{eqnarray}
One can find the wave is circularly polarized at $k=0$ where
$|E_x|=|E_y|$, and as $k$ increases, the wave is elliptically
polarized. The upper and lower branches have opposite helicity, and
$|E_x|>|E_y|$ for upper branch and $|E_x|<|E_y|$ for lower branch.
One can now calculate the current accompanied with the electric field
$\vec{E}$, which can be obtained from Ohm's law and Eq.~\eqref{eq:16}
\begin{eqnarray}
\label{eq:27}
\begin{pmatrix}
 J_x\\ J_y
\end{pmatrix}
=\sigma(\omega) \begin{pmatrix}
 E_x \\ E_y
\end{pmatrix}
= \frac{\omega \epsilon_d}{4\pi i} \begin{pmatrix}
 \left( \frac{k^2c^2}{\mu_d\epsilon_d\omega^2}-1 \right)E_x \\
-E_y
\end{pmatrix}.
\end{eqnarray}
Due to the spin-orbit coupling, the ratio between $S_y$ and $S_x$ can
then be written as
\begin{eqnarray}
\label{eq:28}
\frac{S_y}{S_x} = - \frac{J_x}{J_y} = \left(
  \frac{k^2c^2}{\epsilon_d\mu_d\omega^2}-1 \right) \frac{E_x}{E_y}
\end{eqnarray}
We plot $S_y/(iS_x)$ as a function of $k$ in the right panel of
Fig.~\ref{fig:5}, which is a real number. This indicates the spin
orientation in space is rotating circularly at $k=0$, and elliptically
when $k>0$. This is similar to the two dimensional case as shown in
section II, the only difference is that we have two elliptically
polarized wave here instead of one. Our calculation shows that one may
create the \emph{spin wave plasmon} in stacked HMs by light with
appropriate frequency and incidence along the $xy$ plane.
\section{Conclusion}
In conclusion, the plasmon excitation and the electromagnetic response
of a single layer and multilayers of HMs are carefully investigated for
the case with magnetizations. We find that the ``spin-plasmon'' modes
discussed in Ref.\cite{raghu2010} are modified, with the corresponding
electric fields, as well as the spin orientation, becoming elliptical
due to the Hall response.  Since a single layer of HMs can not strongly
couple to a 3D electromagnetic wave, we consider the electromagnetic
response of the multilayers of HMs stacked by TIs and normal
insulators. For electromagnetic waves incident normal to the HM plane,
we find different plasmon frequencies for the left-hand and right-hand
circularly polarized waves. Therefore, for the light with the frequency
between these two plasmon frequencies, only one type of circularly
polarized wave can propagate along the sample.  For the frequency above
both the plasmon frequencies, a giant Faraday rotation is expected.  For
the light propagating in the HM plane, a new branch of mode appears
below the conventional plasmon frequency due to the mixing between
longitudinal and transverse modes.  Interestingly, the polarization of
these two modes are both elliptical in the helical metal plane, no
longer perpendicular to the wavevector $\vec{k}$. The new optical modes
are expected to be observed in an optical transmission and reflection
experiment\cite{shuvaev2011,li2012,jenkins2012}.
\section{ACKNOWLEDGMENT}
We would like to thank X.L. Qi for the useful discussion. 
This work is supported by NSFC Grant No. 10904081.

\appendix
\section{The retardation effect in the two dimensional plasmon
  excitation of helical metals}
In this appendix, we consider the retardation effect, which has been
neglected for simplicity in Eqs.~\eqref{eq:21} and \eqref{eq:22} in the
text. For this aim, one needs to solve the full Maxwell equations
subject to the time-varying sources. It is then convenient to use the
potentials $\varphi$ and $\vec{A}$, which satisfy the
inhomogeneous wave equations  
\begin{eqnarray}
\label{eq:33}
\left[ \frac{1}{c'^2} \frac{\partial^2}{\partial t^2}-\vec{\nabla}^2\right]
\varphi(x,y,z,t) &=& \frac{4\pi}{\epsilon_d}\rho(x,y,t)\delta(z) \\
\label{eq:34}
\left[ \frac{1}{c'^2} \frac{\partial^2}{\partial t^2} -\vec{\nabla}^2\right]
\vec{A}(x,y,z,t) &=& \frac{4\pi\mu_d}{c}\vec{J}(x,y,t)\delta(z)
\end{eqnarray}
where $c'\equiv c/\sqrt{\mu_d\epsilon_d}$, and the Lorenz gauge in dielectric
material $\vec{\nabla}\cdot
\vec{A}+\mu_d\epsilon_dc^{-1}\partial_t\varphi=0$ is adopted. Since the
system is translation invariant in the $xy$ plane, both
Eq.~\eqref{eq:33} and \eqref{eq:34} can be written in the following form
after Fourier transformation
\begin{eqnarray}
\label{eq:36}
\left[ k^2-\partial_z^2-\frac{\omega^2}{c'^2}
\right]X(\vec{k},z,\omega)
=f(\vec{k},\omega)\delta(z),
\end{eqnarray}
where $\vec{k}$ is the wavevector in the $xy$ plane, and $X$ and $f$ are
the potential and source respectively.  This equation has extended
solutions propagating in the full space with the general form
\begin{eqnarray}
\label{eq:37}
X(\vec{k},z,\omega) = \alpha e^{ik_zz}+\beta e^{-ik_zz}-
\frac{f(\vec{k},\omega)}{2k_z}\sin (k_z|z|)
\end{eqnarray}
where $k_z\equiv\sqrt{(\omega/c')^2-\vec{k}^2}$, and $\alpha$ and
$\beta$ are arbitrary numbers. There is also a localized solution
propagating only in the $x-y$ plane which decays exponentially along the
$z$ direction,
\begin{eqnarray}
\label{eq:38}
X(\vec{k},z,\omega) = \frac{f(\vec{k},\omega)}{2\lambda}e^{-\lambda|z|},
\end{eqnarray}
where $\lambda=\sqrt{\vec{k}^2-(\omega/c')^2}$. In this paper we are
only interested in the localized solution Eq.~\eqref{eq:38}. By
substituting $X$ with $\varphi$ and $\vec{A}$, and $f$ with $\rho$ and
$\vec{J}$, we have
\begin{eqnarray}
\label{eq:39}
\varphi(\vec{k},z,\omega) &=&
\frac{2\pi\rho(\vec{k},\omega)}{\epsilon_d\lambda} e^{-\lambda|z|}, \nonumber\\
\vec{A}(\vec{k},z,\omega)&=&
\frac{2\pi\mu_d\vec{J}(\vec{k},\omega)}{c\lambda} e^{-\lambda|z|}
\end{eqnarray}
Then the \emph{inplane} electric
field($\vec{E}\equiv-\vec{\nabla}\varphi-c^{-1}\partial_t\vec{A}$) reads
\begin{eqnarray}
\label{eq:40}
\vec{E}(\vec{k},z=0,\omega) &=&
 \frac{2\pi\vec{k}}{i\epsilon_d\lambda} \rho(\vec{k},\omega)
+\frac{2\pi i\omega\mu_d}{c^2\lambda}\vec{J}(\vec{k},\omega)
\end{eqnarray}
Now if one assumes $\vec{k}=(0,k,0)$ along the $y$ axis, keeping the
lowest order terms expanded in terms of $c^{-2}$, and using the
continuity condition for the currents, then one can obtain
Eqs.~\eqref{eq:21} and \eqref{eq:22}. 

The group velocity in the nonmagnetized case can be calculated to be
proportional to $k^{1/2}$ as $k\rightarrow0$ without considering the
effect of retardation. It is then obvious that $v_g\propto k^{-1/2}$
which is divergent as $k\rightarrow0$. This unphysical feature can be
cured by taking the retardation effect, i.e., by using
Eq.~\eqref{eq:40}. In the nonmagnetized case, the longitudinal and
transverse fields are decoupled. Eq.~\eqref{eq:40} indicates that the
longitudinal electric field is determined by the longitudinal current
\begin{eqnarray}
\label{eq:43}
E_y(\vec{k},0,\omega) = \frac{2\pi\lambda J_y}{i\epsilon_d\omega} .
\end{eqnarray}
According to the Ohm's law $J_y={i\chi_0}E_y/{\omega}$, then we
require
\begin{eqnarray}
\label{eq:44}
\lambda=\frac{\epsilon_d\omega^2}{2\pi\chi_0(\omega)}
\end{eqnarray}
For small $\omega$, we can approximate $\chi_0(\omega)\approx
\alpha^2\epsilon_d/(2\pi)$ with $\alpha$ defined in
Sec.~\ref{sec:spin-plasmon-modes}. Then
\begin{eqnarray}
\label{eq:45}
\left( \frac{\omega}{\alpha} \right)^2 = \frac{2k^2}{\left( \frac{\alpha}{c'} \right)^2+\sqrt{\left( \frac{\alpha}{c'} \right)^4+4k^2}}
\end{eqnarray}
If $k\gg \alpha/c'$, we recover the plasmon frequency $\alpha \sqrt{k}$
in Ref.\cite{raghu2010}. As $k\rightarrow0$ we can not neglect the
effect of retardation anymore, in fact, in the opposite limit $k\ll
\alpha/c'$, we obtain $\omega\approx c'k$.

\end{document}